\documentclass[fleqn,10pt]{wlscirep}
\title{Polarization tunable all-dielectric color filter based on cross-shaped Si nanoantennas}

\author[1,*]{Vishal Vashistha}
\author[2,]{Gayatri Vaidya}
\author[1]{Pawel Gruszecki}
\author[1]{A E. Serebryannikov}
\author[1,*]{Maciej Krawczyk}
\affil[1]{Faculty of Physics, Adam Mickiewicz University in Poznan, Poland.}
\affil[2]{Centre of Excellence in Nanoelectronics - CEN, IIT Bombay, India.}
\affil[*]{visvas@amu.edu.pl}
\affil[*]{krawczyk@amu.edu.pl}

\usepackage[mediumspace,mediumqspace,squaren]{SIunits}
\usepackage{amsmath}
\usepackage{caption}
\usepackage{subcaption}

\begin{abstract}
Color filters have important applications in the area of Nano-spectroscopy and ccd imaging applications. Metallic nanostructures provide an efficient way to design and engineer ultrathin color filters. 
These nanostructures have capability to split the white light into fundamental colors and enable color filters with ultrahigh resolution but their efficiency can be restricted due to high losses in metals especially at the visible wavelengths. 
In this work, we demonstrate Si nanoantennas based all-dielectric color filters, which are sensitive to incident-wave polarization and, thus, tunable with the aid of polarization angle variation. Two different information can be encoded in two different polarization states in a single physical nanostructure. The nanoantenna based pixels are highly efficient and can provide high quality of colors due to low losses in dielectric at optical frequencies. We experimentally demonstrate that a variety of colors can be achieved by changing the physical size of the nonsymmetric cross-shaped nanoantennas. The proposed devices cover an extended gamut of colors on CIE-1931 chromaticity diagram due to the existence of high quality resonance in Si nanoantennas. The device shows significant tunability of color while operating this color filter device in transmission as well as in reflection mode.
\end{abstract}

\begin{document}

\flushbottom
\maketitle
%
%
\thispagestyle{empty}

\section*{Introduction}
Plasmonic nanostructures suggest a promising platform for replacement of organic dyes printing in future~\cite{kristensen2016plasmonic,brongersma2015introductory,gu2015color,pendry2006photonics}. The normal organic dyes based inclusions split the white light into fundamental color components, i.e., red, green and blue. However, the resolution of these dyes is not sufficient because of large thickness of the polymer that is needed to absorb a certain portion of the white light. Moreover, it has poor resolution which is degradable with time. In addition, these polymers are not eco-friendly. The new era of designing color filters is connected with the concept of light-matter interaction that allows to engineer color filters having several important advantages over organic dyes based printing technology~\cite{kim2009solid}. These advantages include ultra-small size, low power consumption, everlasting colors, and high resolution wide color gamut even below the diffraction limit~\cite{clausen2014plasmonic,tan2014plasmonic,james2016plasmonic,xu2010plasmonic}. In fact, the first ideas regarding interaction of light with a nanostructure have been developed in the ancient time. In particular, the Roman Lycurgus Cup should be mentioned~\cite{freestone2007lycurgus}. Simple examples of light-matter interaction can be observed in nature such as color of sky, wings of butterflies, beetles, and the feathers of peacocks~\cite{srinivasarao1999nano,vukusic2001structural}. The light-matter interaction depends on size and shape of the object. Generally, operation of a plasmonic color filter is based on separation of the white light components with the aid of the engineered plasmon resonance. The resonance characteristics can be statically tuned by altering the physical size of the nanostructure and its composition, which in turn determine wavelengths of light that are scattered, absorbed, or transmitted~\cite{hutter2004exploitation}. But what is most interesting and important for practical applications is that it also depends on the angle of incidence and polarization of incident light. These dependencies can be efficiently utilized in design of dynamically tunable color filters that do not need biasing. Several polarization sensitive color filters have earlier been reported, which operate in transmission mode~\cite{li2016dual,shrestha2015polarization,zeng2014ultrathin,ellenbogen2012chromatic}. Most of the known performances are realized using plasmonic (metal) nanostructures. Earlier the choice for realization of these nanostructure was gold and silver. However, gold is costly for large scale fabrication. It also suffer interband transitions in the visible frequency range, while silver faces the problem of aging, since it is highly reactive with native oxides~\cite{ehrenreich1962optical}. Aluminum might be a good choice because of high stability and low cost~\cite{desantis2016walking,diest2013aluminum,knight2013aluminum}. Unfortunately, Al is even more lossy than gold and silver, especially near 800 nm where it shows interband transition. Ultimately, plasmonic color filters have high losses at the visible wavelength regime. Most of the earlier suggested filters operate in transmission mode. Few of them operate in the dual mode that involves reflection and transmission modes~\cite{park2016trans,yu2014transmissive}. Recently, the investigations have been conducted with the aim to design and fabricate all-dielectric based color filters~\cite{proust2016all,zhao2016full,hojlund2014angle} and hybrid ones which combine metallic and dielectric components~\cite{shrestha2015polarization}. Another methodology is investigated with the use of complementary design methods with the hope of high quality saturated colors~\cite{lee2017wide} which can cover a wide gamut on CIE-1931 chromaticity diagram.  However, operation of these filters is polarization insensitive.

All-dielectric, metasurface-based devices are considered to be promising due to significant advantages over metallic nanostructures such as high quality resonances and low intrinsic ohmic losses~\cite{jahani2016all,decker2015high,sautter2015active,li2016all,chong2015polarization,bonod2015silicon}. The advantages of dielectric, e.g., Si nanodisks includes high refractive index and ease of integration within the well established CMOS technology. The high refractive index of Si allows one to properly manipulate magnetic and electric components of light simultaneously, while in case of metal nanoantennas absorption losses are high and interaction with magnetic component of light is poor. Recently, all-dielectric structures based on Si nanoresonators have been suggested for creating low-loss high quality color filters~\cite{proust2016all,vashistha2016all}. All dielectric nanoantennas based colors filters enables the high quality of colors with extended gamut~\cite{vashistha2016all}. These color filters operate in reflection mode and show high quality of the colors. The high quality saturated colors using Si nanoantenna are due to the inherent property of high quality resonances (sharp resonances compared to metal nanoantenna). This results in wide gamut of saturated colors~\cite{vashistha2016all}. The recent progress in the area of light-matter interaction allows to efficiently utilize different spectral components of light in different regimes in one device. In this work, we demonstrate all-dielectric color filters, which are composed of nonsymmetric Si nanoantennas unlike the earlier reported using metal based plasmonics nanoantennas~\cite{li2016dual,shrestha2015polarization,zeng2014ultrathin,ellenbogen2012chromatic}. These non-symmetric si nanoantenna based color filters are strongly sensitive to the incident-wave polarization and, thus, can be tuned by changing polarization. The proposed nanoantennas show low losses at the visible spectrum and high confinement of light. We experimentally demonstrate that the suggested color filters can operate in transmission and reflection modes and can be tuned by changing the incident-wave polarization. The low losses of Si nanoantennas result in high quality of colors in both modes. The designed nanoantennas are asymmetric cross-shaped, so that the resonances depend on polarization, as desired. Thus, two information can be encoded in one physical structure, and then decoded by using vertically and horizontally polarized wave. In contrast with the recently reported performances of color filters, the ones suggested in this paper combine advantages of all-dielectric design, with wide gamut high saturated colors in dual mode operation and sensitivity to polarization.

\section*{Results}
\textbf{Operation Principles}
First, let us consider a single rectangular Si nanoantenna. Its width, length and height were chosen as 40{\nano\meter}, 60{\nano\meter}, and 200{\nano\meter}, respectively. The nanoantenna is excited by a horizontally polarized wave in the visible range. Since spectral location of resonance is a function of nanoantenna length, we have scaled it by varying the length from 60{\nano\meter} to 200{\nano\meter}. As expected, spectral location of the resonance is shifted from lower visible spectrum to higher visible spectrum.

\begin{figure}[!htb]
\centering
\includegraphics[width=12cm,height=8cm]{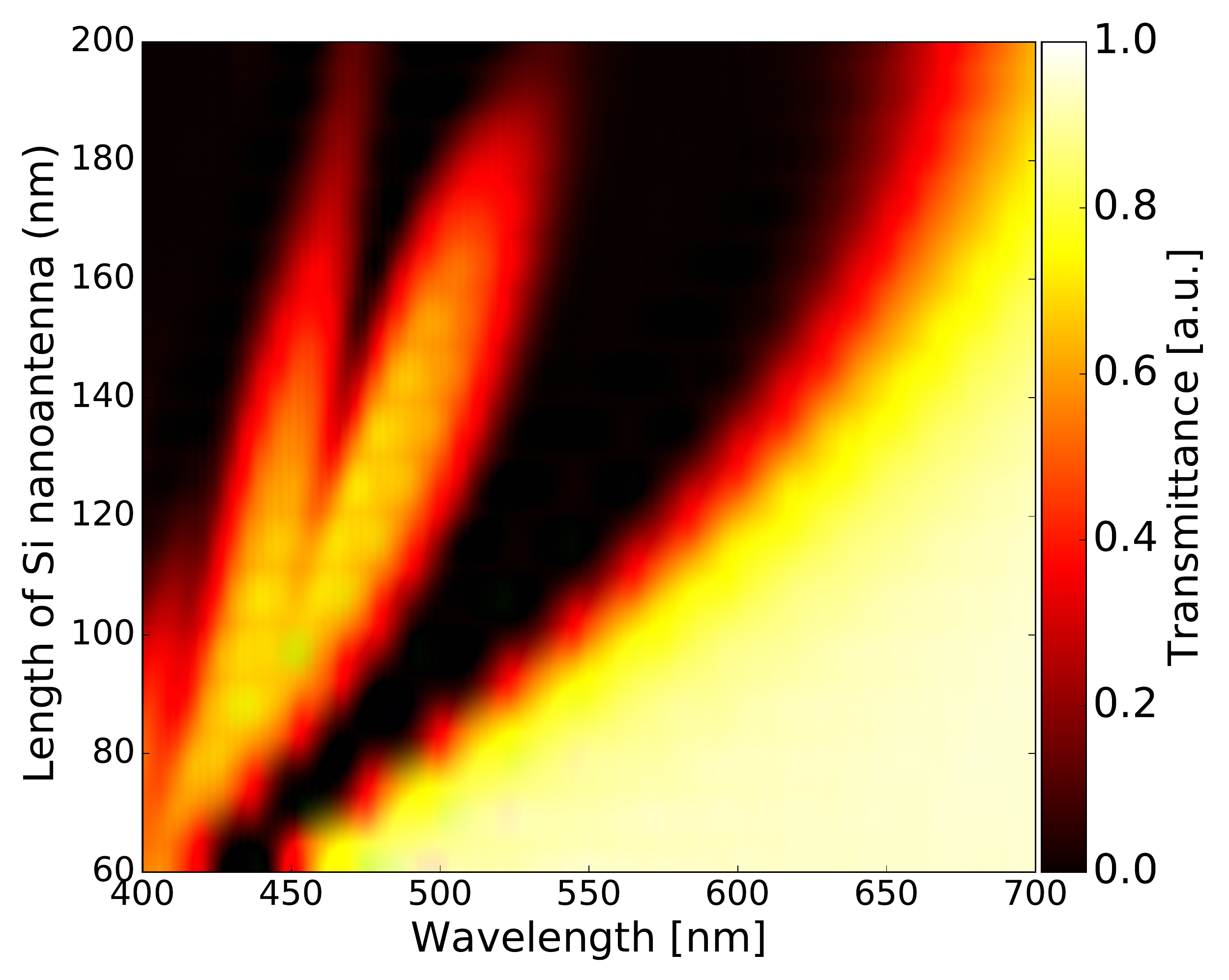}
\caption{(a) Transmittance of Si rectangular nanoantennas when its length is gradually varied from 60{\nano\meter} to 200{\nano\meter}.}
\label{length_sweep}
\end{figure}

\begin{figure*}[!htb]
\centering
\includegraphics[scale=1.0]{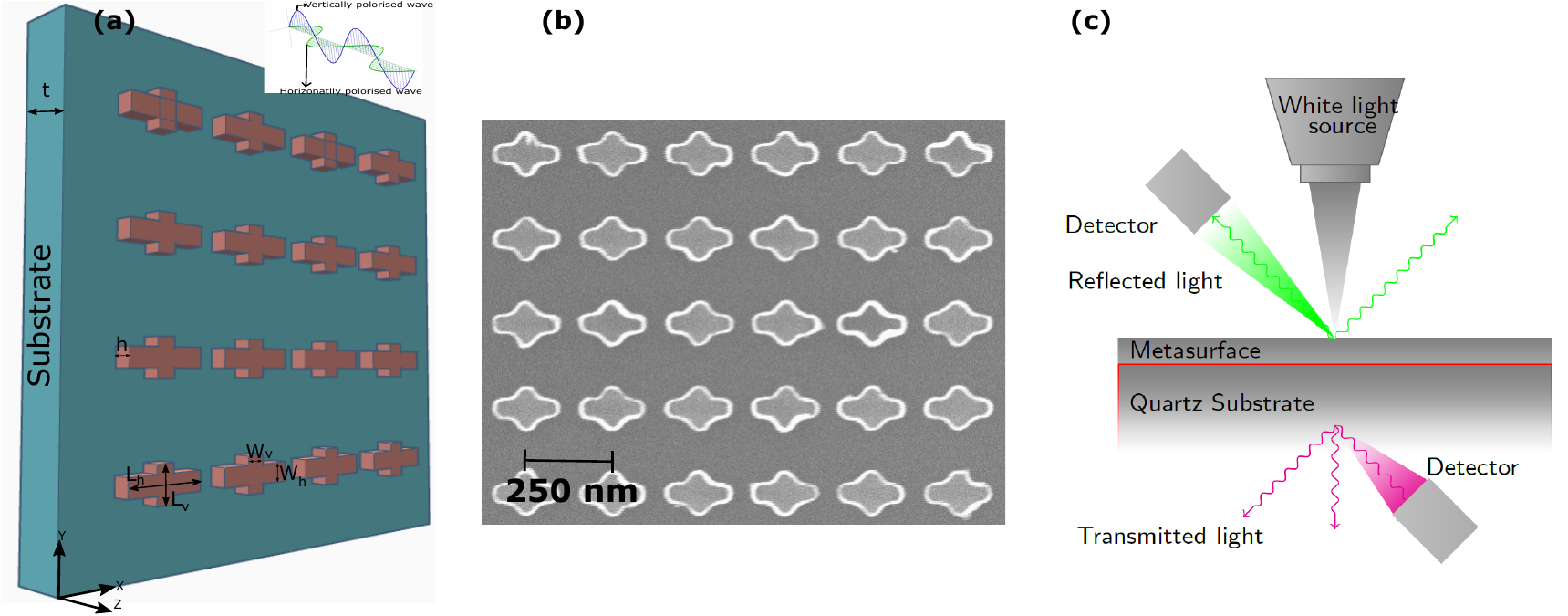}
\caption{(a) Metasurface composed of nonsymmetric Si nanoantennas placed on top of quartz substrate; inset schematically shows propagation in case of horizontally $\phi=0^{\circ}$ and the vertically $\phi=90^{\circ}$ polarization of incident wave. (b) Top view of SEM image of fabricated device. (c) A schematic representation of the color filter operating in transmission and reflection modes.}
\label{main_figure}
\end{figure*}

Figure~\ref{length_sweep} presents the simulated results for shift in transmittance dip towards larger wavelengths while increasing the length of nanoantenna. The extinction cross section of the nanoantenna also shifts towards larger wavelengths with the increase in length (see more details in supplementary information, Figure S1). From Fig.~\ref{length_sweep}, it is clear that any arbitrary color can be achieved by properly selecting the appropriate length of the rectangular nanoantenna. Based on these results illustrated in Fig.~\ref{length_sweep}, we have selected two rectangular rod nanoantennas and combined them to form a nonsymmetric cross-shaped geometry. Freedom in choice of the length ratio of the longer to the shorter segment of individual cross-shaped nanoantennas allows us to obtain resonance frequencies which depend on polarization state of the incident wave. Thus, operation of the resulting nanoantenna is polarization sensitive. The general geometry of the designed metasurface is shown in Fig.~\ref{main_figure}. The lengths of the horizontal (longer) ($L_h$) and vertical (shorter) ($L_v$) segments of each nonsymmetric nanoantenna are 150{\nano\meter} and 90{\nano\meter}, respectively, while the width ($W=W_h=W_v$) and height ($h$) are 40{\nano\meter} and 200{\nano\meter} for both segments. The lattice constant of the structure is 250{\nano\meter}. The Si nanoantennas are placed on top of the quartz substrate whose thickness is $275 \pm 5{\micro\meter}$, see Fig.~\ref{main_figure}(a). 
A schematic illustrating the structure excitation by using the horizontally $(\phi=0^{\circ})$ and the vertically $(\phi=90^{\circ})$ polarized incident waves is shown in Fig.~\ref{main_figure}(a), inset. 
Figure~\ref{main_figure}(b) presents the SEM image of the fabricated device. Figure~\ref{main_figure}(c) presents the schematic of the device operating in transmission mode and reflection mode. 

Response at each of two orthogonal polarization states is mainly determined by the length of either the longer or the shorter segment. For arbitrary polarized light, the combined response is given by the equation.
\begin{equation}
\centering
T (\phi,\lambda) = T_v(\lambda)\sin^2\phi + T_h(\lambda)\cos^2\phi,
\label{eq1}
\end{equation}
where \(\phi\) is polarization angle of the incident wave vector and \(\lambda\) is the wavelength in free space. \(T_v(\lambda)\) and \(T_h(\lambda)\) mean transmittance for vertically and horizontally polarized light, respectively. \(T (\phi,\lambda)\) is the total transmittance due to all polarization components. 

\begin{figure}[!ht]
\centering
\hspace{-2.2em}
\includegraphics[scale=0.6]{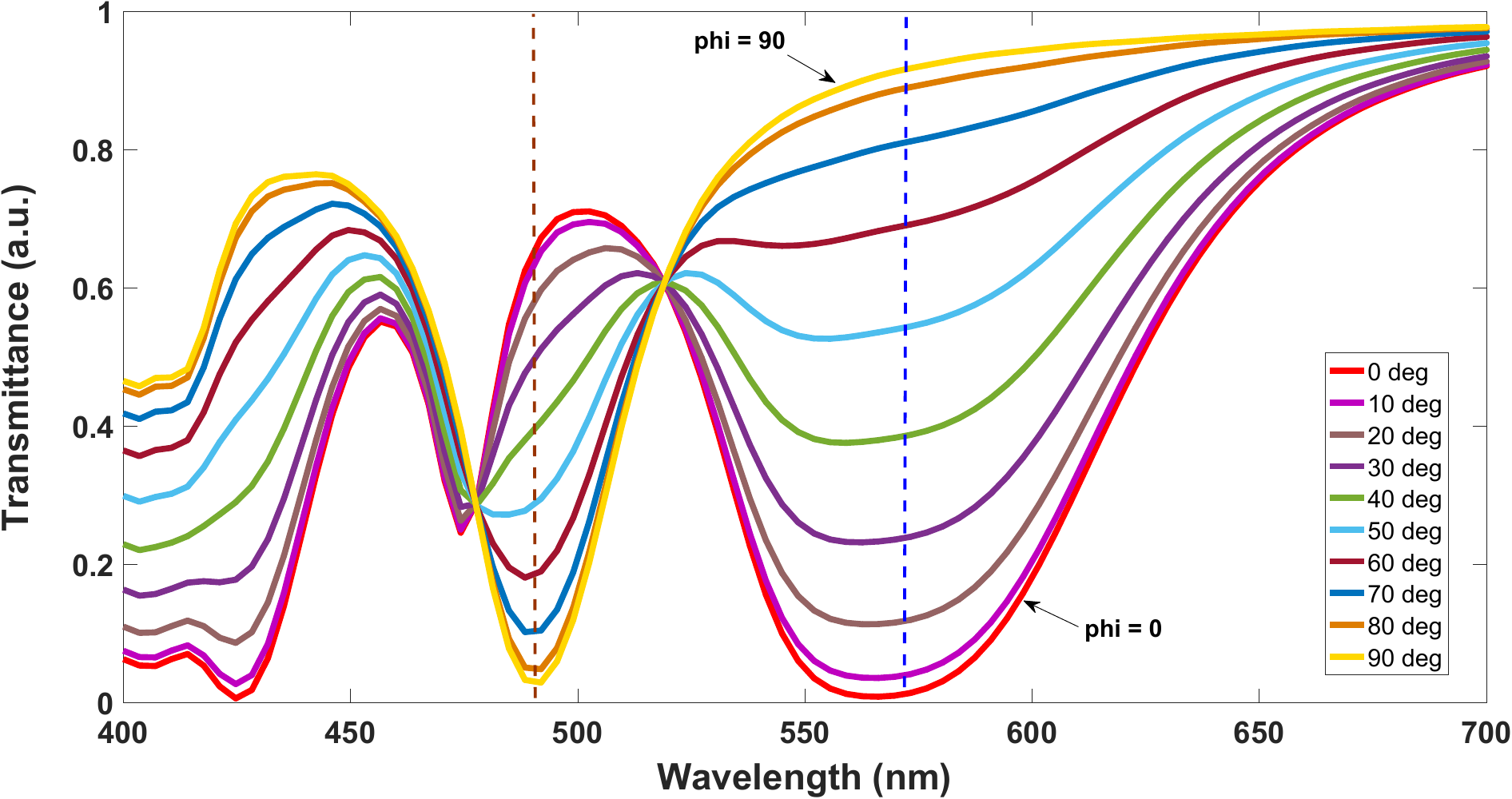}
\caption{Shift in transmittance spectra when polarization of incoming wave is changed from 0$^{\circ}$ to 90$^{\circ}$. Dashed lines indicate two resonance cases with strong transmission contrast between two orthogonal polarization states.}
\label{polarization_1}
\end{figure}

In order to illustrate sensitivity of the combined response of the device to variations of polarization, we performed simulations by changing the value of $\phi$ with a step of $10^{\circ}$. The results are presented in Fig.~\ref{polarization_1}. One can see that strong redistribution of the incident-wave energy between the transmitted and reflected waves takes place in the whole wavelength range considered (see supplementary information, Fig. S2 to S5 ). Moreover, there is strong sensitivity to the polarization state. For instance, the lowest Mie resonance is expected appears for the longer antenna segment near 570{\nano\meter}, where the transmission dip appears for one of two orthogonal polarization, $\phi=0^{\circ}$, and then gradually disappears while $\phi$ is increased. 
In turn, transmission dip for the second orthogonal polarization ($\phi=90^{\circ}$), which should be connected with the lowest Mie resonance of the shorter segment, appears near 480{\nano\meter}.
The obtained results allow one to expect that different colors can be obtained at different polarization states. As the length of the nanoantenna is increases, a higher-order Mie resonance also appears for the longer segments, which leads to the additional dip at 420{\nano\meter} when $\phi=0^{\circ}$. More details regarding effect of variation of $\phi$ for both is given in supplementary information in Fig. S2 to S5.

\begin{figure}[!htb]
\centering
\includegraphics[scale=0.70]{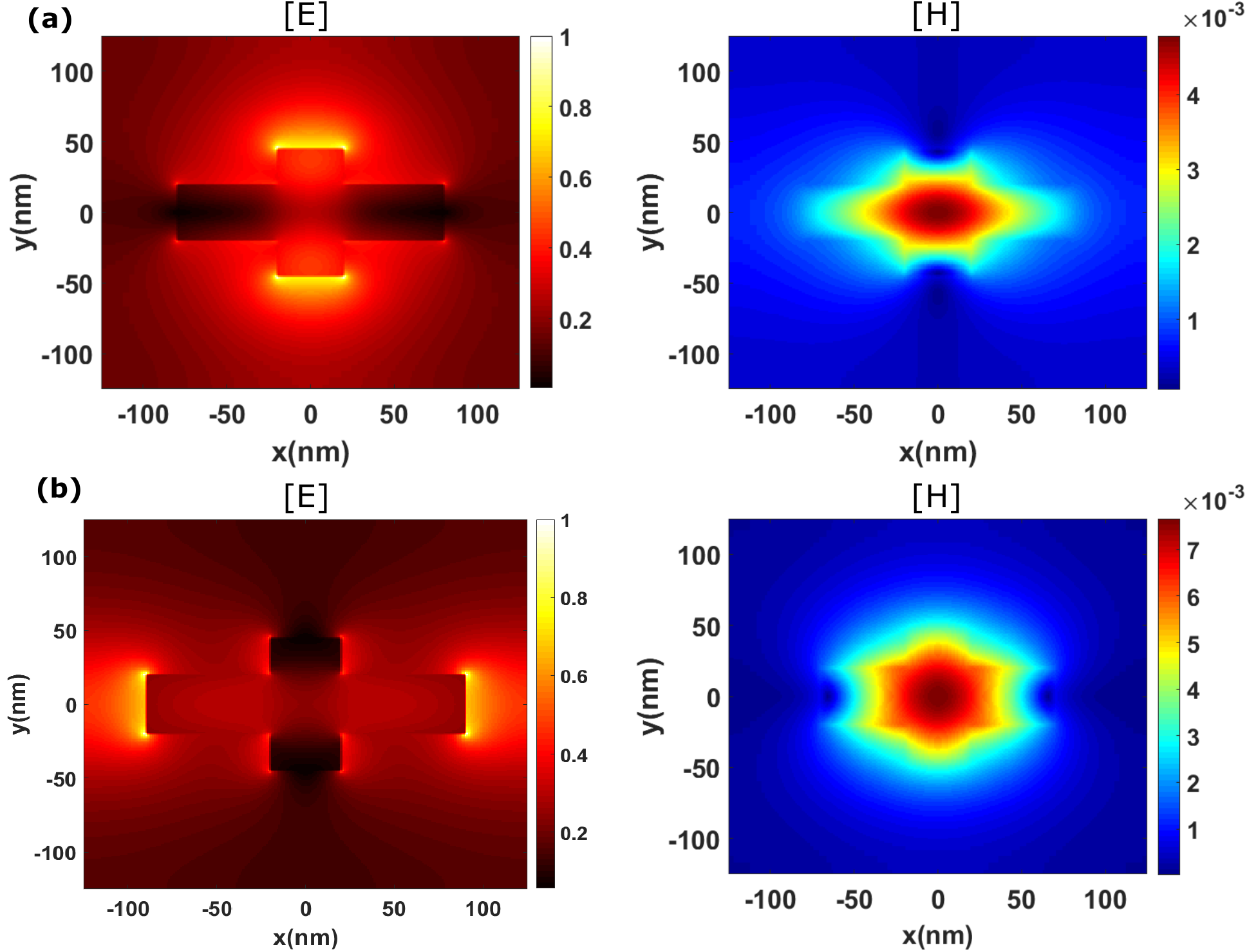}
\caption{Electric and magnetic field pattern for vertically and horizontally polarized wave (a) Electric (left) and magnetic (right) field pattern in $xy$ plane in the middle of the Si nanostructure at 480{\nano\meter} wavelength for vertically polarized wave. (b) Electric (left) and magnetic (right) filed pattern in $xy$ plane of the Si nanostructure at 570{\nano\meter} wavelength for horizontally polarized wave.}
\label{field_plot}
\end{figure}

To further clarify the operation principles of individual nanoantenna as a nanopixel, we have drawn the field pattern of the cross-shaped nanostructure for two orthogonal states of polarization of the incoming wave. Each nanoantenna i.e., vertical and horizontal responds individually when electric vector of incident wave align with the geometry. In Fig.~\ref{field_plot}, electric and magnetic fields are plotted for $\phi= 0^{\circ}$ and 90$^{\circ}$. When the electric field vector of incoming wave is align with horizontal or vertical nanoantenna, the corresponding segment behaves like an individual pixel, whose response can be tuned by varying the length. Thus, when the electric field is perpendicular, i.e., $\phi=90^{\circ}$, only the vertical nanoantenna responds, that justifies the resonance observed in Fig.~\ref{polarization_1} near 480{\nano\meter}. For $\phi= 0^{\circ}$, horizontal nanoantenna is in excitation mode, and this results in resonance 
observed in Fig.~\ref{polarization_1} near 570{\nano\meter}. The magnetic field is confined around the center of the cross in both cases, so strong difference in electric field is the origin of polarization sensitivity. 

\begin{figure}[!htb]
\centering
\includegraphics[scale=0.7]{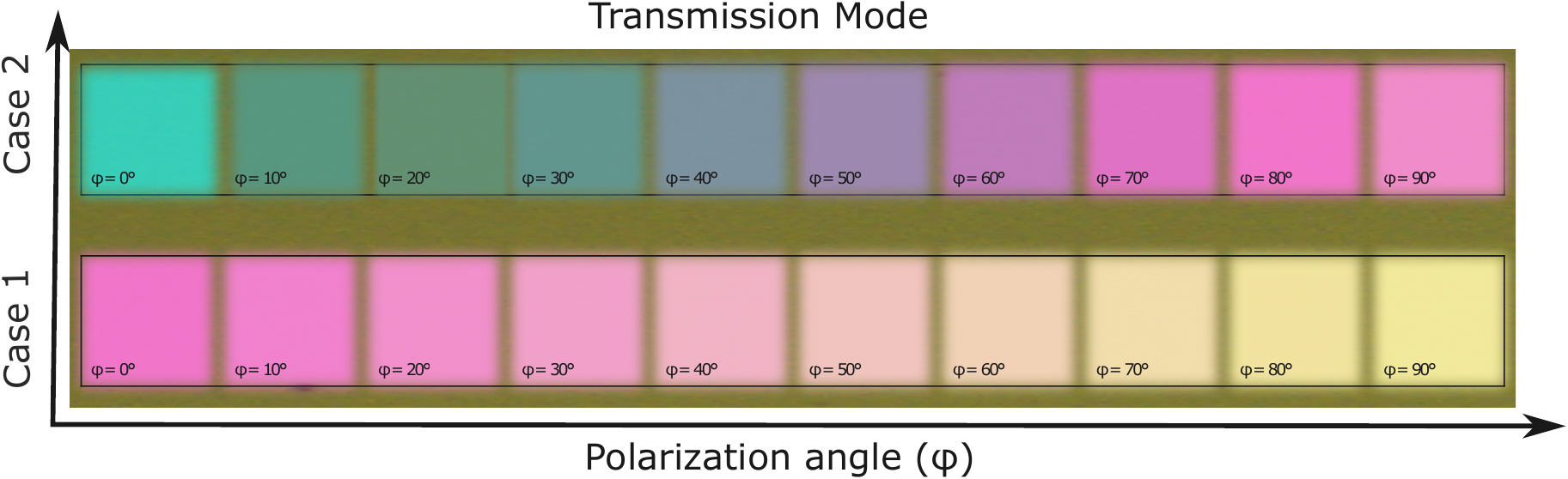}
\caption{Colors visible under optical microscope in transmission mode when polarization angle is gradually varied from($\phi=0^{\circ}$ to $\phi=90^{\circ}$). Dimensions of nanoantennas are given in Table~\ref{dimensions}.}
\label{optical_trans}
\end{figure}

\textbf{Experimental Validation}
Since white light contains all colors, so its each spectral component is associated with a certain color in transmission and reflection modes. We have considered two nanoantenna design cases with two different dimension sets which are given in Table~\ref{dimensions}. 
These two cases are chosen so that we could demonstrate the primary colors, i.e., RGB or CMY.
Indeed, we observed different colors for different states of polarization of the incoming wave in transmission mode. Figure~\ref{optical_trans} shows the colors observed under the optical microscope in transmission mode when the polarization angle is gradually varied from($\phi=0^{\circ}$ to $\phi=90^{\circ}$). 

\begin{table}[htbp]
\centering
\caption{\bf Dimensions of the studied nanoantennas in nm.}
\begin{tabular}{cccccc}
\hline
       & $L_v$  & $L_h$  & $W_v$ & $W_h$ & $h$   \\
\hline       
Case 1 & 90  & 160 & 40 & 40 & 200 \\
Case 2 & 110 & 190 & 40 & 40 & 200 \\
\hline
\end{tabular}
\label{dimensions}
\end{table} .  

\begin{figure}[!htb]
\centering
\includegraphics[scale=0.7]{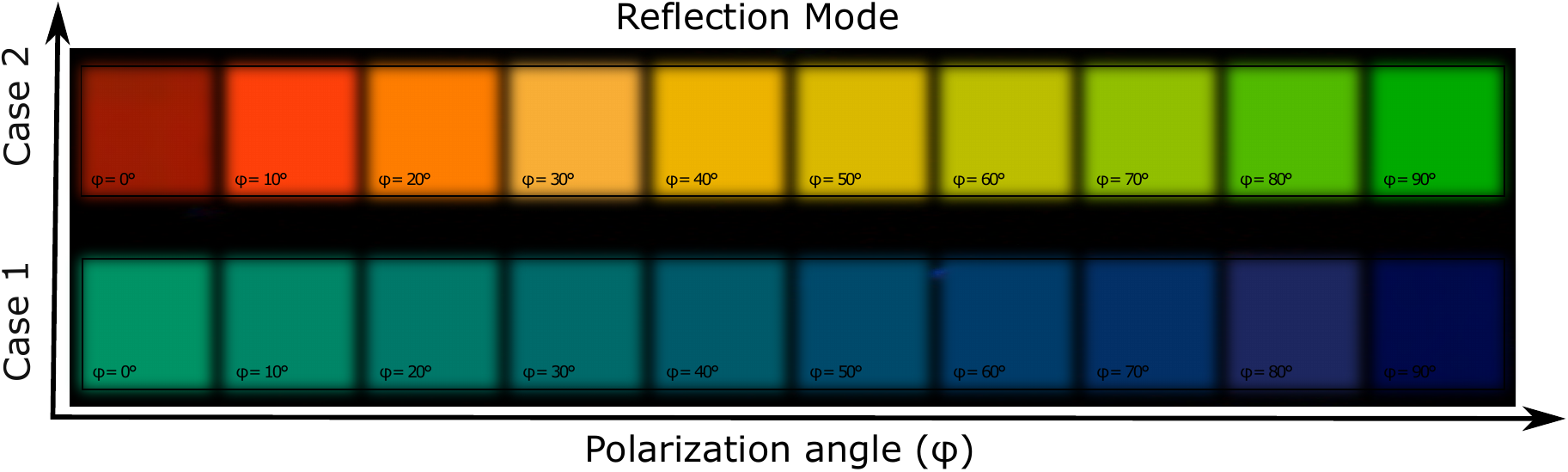}
\caption{Colors visible under optical microscope in reflection mode when polarization angle is gradually varied from($\phi=0^{\circ}$ to $\phi=90^{\circ}$). Dimensions of nanoantennas are given in Table~\ref{dimensions}.}
\label{optical_ref}
\end{figure}

Whilst location of the dip in transmission is function of nanoantenna length, the transmission dip can be shifted within a large part of the visible spectrum by scaling nanoantennas. So, we can achieve a wide variety of colors to encode the two different information by simply adjusting the lengths of the vertical and horizontal nanoantenna segments. A specific portion of the incident white light is reflected from the metasurface, while another part is transmitted through the quartz substrate, as schematically shown in Fig.~\ref{main_figure}(c). The dielectric nanoantennas have very low losses that enables creation of different colors in transmission mode as well as reflection mode. The cross shape of nanostructure elements allows splitting the light into its spectral components depending on the size of the nanoantenna segments. When the same device works in reflection mode, we obtain different colors for different states of polarization of the incoming wave, so tunability can be realized by means of polarization modification. Figure~\ref{optical_ref} shows the colors seen under optical microscope in reflection mode for each of two orthogonal polarization.

\begin{figure*}[!ht]
  \centering
  \includegraphics[scale=0.9]{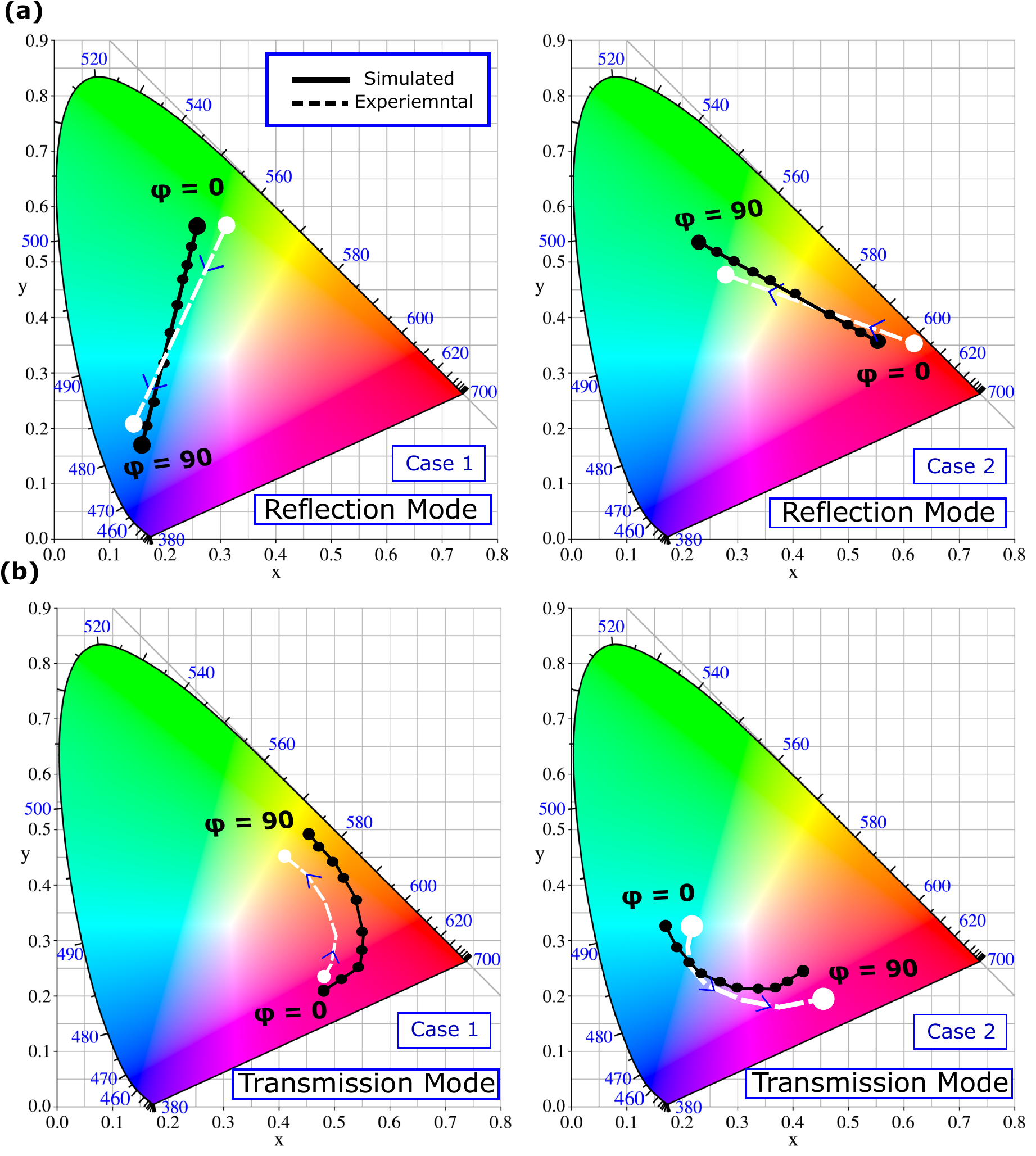}
 \caption{Polarization related changes in color: (a) for reflection mode, from green to blue in case 1 and from red to green in case 2; (b) for transmission mode, from magenta to yellow to in case 1 and from cyan to magenta in case 2; polarization of incident white light is changed for experimental results from horizontal $(\phi=0^{\circ})$ to vertical $(\phi=90^{\circ})$; for comparison, simulation results for polarization states from $\phi=0^{\circ}$ to $\phi=90^{\circ}$ with the step size of $10^{\circ}$ are also shown.}
\label{trans_ref}
\end{figure*}

A dual characterization is carried out to justify the above discussed results. We have used a home-made customized setup to characterize operation of the device. The details of the optical characterization are available in supplementary information. The experimental transmitted and reflection spectra are converted into the colors by using standard chromaticity matching functions~\cite{smith1931cie}, see supplementary information. For the 
cases 1 and 2, the gradual color changes are observed when polarization is gradually varied from horizontal $(\phi=0^{\circ})$ to vertical one $(\phi=90^{\circ})$. 
In Fig.~\ref{trans_ref}, 
the results are plotted on the standard CIE 1931 chromaticity diagram. Besides, the experimental results are compared to the simulated ones.

In fact, there is some discrepancy between simulation and experimental results in Fig.~\ref{trans_ref}, which 
appears due to imperfections in fabrication and limitations in measurement accuracy. In particular, it can be difficult to fabricate a nanostructure with a high-aspect ratio. For case 1 in reflection mode, it is expected from the simulation results that the changes in color should occur from green to blue zone when polarization of incident light is changed from 0$^{\circ}$ to $90^{\circ}$, and so happens in the experimental results with acceptable deviation. And similar acceptable changes are observed for case 1 in transmission mode, where color changes are predicted from magenta to yellow zone, based on the simulation results, and so happens in the experimental results. A similar level of results coincidence has also been found for case 2, in both reflection and transmission modes. Hence, there is a good trade off between the simulation and experimental results, as far as the latter are still in the predicted zone of the colors.

\section*{Discussion}
To summarize, we have studied all-dielectric metasurfaces based on nonsymmetric cross-shaped Si nanoantennas that are designed for filtering colors in reflection and transmission modes simultaneously. The proposed low-loss, Si-based, all-dielectric color filters enable high quality of colors for the both modes. The resonance location can be adjusted by properly selecting the length and width of the nanoantennas. The nonsymmetric shape of an individual nanoantenna makes it sensitive to the polarization state of incident wave. Hence, efficient tuning of colors can be achieved by changing polarization. A further high accuracy results can be obtained by improvement in fabrication quality and measurement setup. The proposed device cover a wide gamut of colors on CIE-1931 chromaticity diagram. The suggested devices have potential applications in the area of secured optical tag, nano spectroscopy, fluorescence microscopy or CCD imaging.

\section*{Methods}

\textbf{Simulations.} We have used Lumerical FDTD solver to study the cross-shaped nanoantennas on quartz substrate. The materials used for substrate and cross-shaped nano antenna are $SiO_2$ (glass)-Palik and $Si$ (Palik), respectively. These material parameters are taken from default material library of the used software. A plane wave ranging the wavelength from 400{\nano\meter} to 700{\nano\meter} is illuminated from the top of the structure. Periodic boundary conditions are used in the unit cell along $X$ and $Y$ directions. Perfect matching layer (PML) boundary conditions were used in the $Z$ directions to avoid any reflection. The reflectance and transmittance spectra are simulated by considering a unit cell (single cross shaped nanoantenna on substrate) with periodic boundary conditions in $X$ and $Y$ directions.\\\\
\textbf{Device fabrication.} A piranha cleaned quartz sample (275{\micro\meter} thick) is used to fabricate the device. We have deposited a thin layer of 200{\nano\meter} amorphous $Si$ using ICPCVD tool at 300\degree Celsius with 150watt added microwave power. A single layer PMMA photoresist is used for patterning cross shaped nanoantennas by using Raith 150-Two EBL tool. An electronic mask is designed using open source python program. Exposed sample is developed using MIBK-IPA (1:3) and IPA solution for 45{\second} and 15{\second}, respectively. A thin layer of metal (5{\nano\meter} Cr as adhesion layer and 40{\nano\meter} Au ) is deposited to transfer the pattern on metal layer for lift-off process using four target evaporator. After lift-off, the sample is etched using plasma asher to get the final pattern. A process flow chart with step by step details is available in supplementary information. \\\\
\textbf{Optical characterization.} A dual optical characterization is done to ensure the results. The sample is placed under Olympus optical microscope illuminated with white light without filter. The colors can be directly seen under optical microscope in reflection and transmission mode by changing the polarization of the light. The reflectance and transmission spectra is measured using a home made customized setup. A HL 2000 halogen lamp source is coupled with optical fiber to illuminate the light in visible range (400{\nano\meter} to 700{\nano\meter}) onto the sample. A polarizer is added in the path of the optical fibre to control the polarization. A 50$\times$ objective lens is used to tight focus the light on the sample. The reflectance and transmittance spectra is measured using the same objective lens. The data is normalized with respect to bare quartz sample. A Nikon camera attached with the assembly is used to take the photograph of illuminated area.

\bibliography{sample}

\section*{Acknowledgments}

This work was supported by INUP program funded by DeitY (11dit005), Government of India, National Science Centre Poland for OPUS grant No. 2015/17/B/ST3/00118(Metasel) and by the European Union Horizon2020 research and innovation program under the Marie Sklodowska-Curie grant agreement No. 644348 (MagIC). Authors thanks to all members of CEN laboratory, IIT Bombay who helped us directly or indirectly while doing nanofabrication work. Special thanks to Dr.\ K Nageshwari and Dr.\ Ritu Rashmi for providing necessary facilities and regular advice.

\section*{Author contributions statement}

VV is responsible for theoretical simulation and optical characterization of the sample. GV is responsible for optimization of EBL and liftoff process. AS, PG and MK have discussed and analyzed the simulation and fabrication results, and contributed to writing the manuscript. All authors have reviewed the manuscript before submission. 

\section*{Additional information}
\textbf{Supplementary information} accompanies this paper at http://www.nature.com/srep \\\\
\textbf{Competing financial interests:} The authors declare no competing financial interests.

\end{document}